\documentclass[12pt]{article}
\usepackage{epsfig}

\def\beq{\begin{equation}}
\def\eeq#1{\label{#1}\end{equation}}
\def\eeqn{\end{equation}}

\def\beqa{\begin{eqnarray}}
\def\eeqa#1{\label{#1}\end{eqnarray}}
\def\eeqan{\end{eqnarray}}

\let\bar=\overbar

\def\ket#1{\left| {#1} \right\rangle}

\def\O{{\cal O}}

\def\Dslash{\not{\hbox{\kern-4pt $D$}}}
\def\dslash{\not{\hbox{\kern-2pt $\del$}}}

\def\msb{{\bar{\ssstyle M \kern -1pt S}}}

\def\beq{\begin{equation}}
\def\eeq{\end{equation}}
\def\bea{\begin{eqnarray}}
\def\eea{\end{eqnarray}}
\def\nn{\nonumber}
\def\roughly#1{\mathrel{\raise.3ex\hbox
{$#1$\kern-.75em\lower1ex\hbox{$\sim$}}}}

\def\ket#1{\left| #1\right\rangle}
\def\bd{B_d^0}

\def\bs{B_s^0}
\def\bsbar{{\bar B}^0_s}

\def\kbar{{\bar K}^0}

\def\btod{{\bar b} \to {\bar d}}
\def\btos{{\bar b} \to {\bar s}}

\def\Bsdecay{\bs \to J/\psi \phi}
\def\BKstar{B\to\phi K^*}
\def\bstautau{{\bar b} \to  {\bar s} \tau^+ \tau^-}
\def\bscc{{\bar b} \to  {\bar s} c {\bar c}}
\def\fT{f_T}
\def\fL{f_L}
\def\fTfL{f_T/f_L}

\def\Title#1{\begin{center} {\Large {\bf #1} } \end{center}}

\begin{document}

\begin{flushright}
UdeM-GPP-TH-12-210
\end{flushright}

\smallskip

\Title{\boldmath $B$ Physics: Theory Overview\footnote{Talk given at
    {\it Physics at LHC 2012}, University of British Columbia,
    Vancouver, BC, Canada, June 2012}}

\bigskip\bigskip


\begin{raggedright}  

{\it David London\index{London, D.}\\
Physique des Particules, Universit\'e de Montr\'eal\\
C.P. 6128, succ.\ centre-ville\\
Montr\'eal, QC, Canada H3C 3J7}
\bigskip\bigskip
\end{raggedright}

\section{Introduction}

A great deal of work on $B$ physics has been done over the past 15-20
years. At first, it was mostly in the context of the $B$ factories
BaBar and Belle.  This included finding methods for measuring the
standard model (SM) parameters, examining ways of looking for new
physics (NP), analyzing the results, etc.  Unfortunately, most of the
measurements at the $B$ factories agreed with the SM. Although there
were several hints of NP, mostly in $\btos$ transitions, there were no
statistically-significant signals.

CDF and D\O\ then demonstrated that $B$ physics can be done at hadron
colliders. They made a number of measurements involving $\bs$ mesons,
in particular $\bs$-$\bsbar$ mixing.

The LHC will continue this exploration of $B$ physics.  They will
focus mainly on $\bs$ mesons, but may well be able to repeat (and
perhaps improve upon?) some of the measurements made at BaBar and
Belle. As always, the hope is to find a signal of NP. It seems clear
now that very large signals are ruled out. But the LHC may well have
the precision to detect even small deviations from the SM. In this
talk, I will discuss a number of $B$-physics measurements to be made
at the LHC that have the potential for revealing NP.

\section{\boldmath $\bs$-$\bsbar$ Mixing}

In the presence of $\bs$-$\bsbar$ mixing, the mass eigenstates $B_{L}$
and $B_{H}$ [$L$ ($H$) corresponds to ``light'' (``heavy'')] are
admixtures of the flavour eigenstates $\bs$ and $\bsbar$:
\bea 
\ket{B_L} &=& p \ket{\bs} + q \ket{\bsbar} ~, \nn\\
\ket{B_H} &=& p \ket{\bs} - q \ket{\bsbar} ~, 
\eea 
with $|p|^2 + |q|^2 =1$. The initial flavour eigenstates oscillate
into one another according to the Schr\"odinger equation with $H = M^s
- i \Gamma^s/2$ ($M^s$ and $\Gamma^s$ are the dispersive and
absorptive parts of the mass matrix), and lead to the time-dependent
states $\bs(t)$ and $\bsbar(t)$. The off-diagonal elements $M^s_{12}$
and $\Gamma^s_{12}$ are generated by $\bs$-$\bsbar$ mixing.

Defining $\Delta M_s \equiv M_H - M_L$ and $\Delta \Gamma_s \equiv
\Gamma_L - \Gamma_H$, we have
\beq 
\Delta M_s =  2 |M_{12}^s| ~,~~
 \Delta \Gamma_s = 2 |\Gamma_{12}^s| \cos \phi_s  ~,~~
 \frac{q}{p} =  e^{-2i\beta_s} ~,
\label{weakphasedefs}
\eeq 
where $\phi_s \equiv \arg(-M_{12}^s/\Gamma_{12}^s)$ is the CP phase in
$\Delta B=2$ transitions. The weak phases $\phi_s$ and $2\beta_s$ are
independent. The SM predicts that both $\phi_s$ and $2 \beta_s$ are
very small (but $\phi_s \ne - 2 \beta_s$!). $\Delta \Gamma_s$ is
sizeable and is positive in the SM. In the presence of NP, one can
have $2\beta_s \ne 0$ and $\Delta \Gamma_s < 0$.

{\bf\boldmath $J/\psi \phi$:} In 2008 CDF and D\O\ measured the
indirect CP asymmetry in $\Bsdecay$, and found a hint for CPV. The
2011 update gives (at 68\% C.L.) \cite{CDFD0update}
\bea
2\beta^{\psi\phi}_s \,&\in& \, \left[2.3^\circ, 59.6^\circ
  \right]\cup\left[123.8^\circ,177.6^\circ \right] ~, 
~~~~~~~~{\hbox{CDF}} ~,  \nn \\
                   &\in& \, \left[ 9.7^\circ , 52.1^\circ
  \right]\cup\left[127.9^\circ , 170.3^\circ \right] ~, 
~~~~~~~~{\hbox{D\O}} ~.
\eea
Note that the measurement is insensitive to the transformation
$(2\beta^{\psi\phi}_s, \Delta\Gamma_s) \leftrightarrow (\pi -
2\beta^{\psi\phi}_s, -\Delta\Gamma_s)$. This implies that
$2\beta^{\psi\phi}_s$ has a twofold ambiguity, which is reflected in
the two ranges of possible solutions above.

LHCb has greatly improved upon this result. First, the twofold
discrete ambiguity has been removed by measuring ${\rm
  sign}(\Delta\Gamma_s)$: $\Delta\Gamma_s = 0.120 \pm 0.028 ~{\rm
  ps}^{-1}$ \cite{signDGs}.  This is done using the decay $\bs \to
J/\psi \phi(\to K^+ K^-)$, and looking at the interference between the
$s$- and $p$-wave $K^+ K^-$ angular momentum states.

Second, they find \cite{betasmeas}
\beq
2\beta_s^{J/\psi \phi} = (-0.06 \pm 5.77~({\rm stat}) \pm 1.54~({\rm
  syst}))^\circ ~,
\eeq
in agreement with the SM. Still, the errors are large enough that NP
cannot be excluded.

To completely search for NP, LHCb has to measure $\bs$-$\bsbar$ mixing
in as many different decays as possible. This has already begun:

{\bf\boldmath $J/\psi f_0(980)$:} LHCb has measured $\beta_s^{J/\psi
  f_0} = (-25.2 \pm 25.2 \pm 1.1)^\circ$ \cite{LHCbpsif0}. The
advantage of this decay is that, because the $f_0(980)$ is a scalar,
no angular analysis is needed. The disadvantage is that, because the
$f_0(980)$ is not a pure $s{\bar s}$ state, there are possibly other
contributions to the decay, leading to hadronic uncertainties
\cite{fleischerpsif0}.

{\bf\boldmath $J/\psi \pi^+\pi^-$:} LHCb has measured $\beta_s^{J/\psi
  \pi^+\pi^-} = (-1.09^{+9.91+0.23}_{-9.97-0.17})^\circ$
\cite{LHCbpsipipi}.  {\it A priori}, since this is a 3-body state, its
CP can be $+$ or $-$, and so it cannot be used to cleanly extract
weak-phase information. However, it has been shown that the $J/\psi
\pi^+\pi^-$ state is almost purely CP $-$ \cite{LHCbpsipipiCP}, so
that there is little error due to the CP $+$ state.

Other final states that are potentially of interest include (i)
{\bf\boldmath $D^{\pm}_s K^{\mp}$} \cite{fleischerDsK} -- here one
extracts $(2\beta_s + \gamma)$, (ii) {\bf\boldmath $D_s^+ D_s^-$}
\cite{fleischerDsDs} -- here one has to deal with penguin pollution,
(iii) {\bf\boldmath $D^0_{CP} K {\bar K}$} \cite{DCPKKbar} -- here one
requires a Dalitz-plot analysis.

Finally, {\bf\boldmath $\bs \to K^{(*)0} {\bar K}^{(*)0}$} is a pure
$\btos$ penguin decay whose amplitude in the SM is $A = V^*_{tb}
V_{ts} P'_{tc} + V^*_{ub} V_{us} P'_{uc}$. The second term is doubly
Cabibbo suppressed with respect to the first term. If it is neglected,
the indirect CP asymmetry vanishes in the SM, so that its measurement
could reveal NP. However, $V^*_{ub} V_{us} P'_{uc}$ is not entirely
negligible. Including it, it is found that indirect CPV measures
$|\beta_s^{eff}| \le 14.9^\circ$ in the SM \cite{BDIL}. If a larger
value of $|\beta_s^{eff}|$ is measured, this would imply NP.

\section{Like-sign Dimuon Asymmetry}

D\O\ has reported an anomalously large CP-violating like-sign dimuon
charge asymmetry in the $B$ system.  The updated measurement is
\cite{D0dimuonnew}
\bea
  A_{\rm sl}^b \equiv \frac{N_b^{++}  - N_b^{--}}{N_b^{++}  + N_b^{--}}
= -(7.87 \pm 1.72 \pm 0.93) \times 10^{-3} ~,
\eea
a 3.9$\sigma$ deviation from the SM prediction, $A_{\rm sl}^{b,{\rm
    SM}} = ( -2.3^{+0.5}_{-0.6} ) \times 10^{-4}$ \cite{Lenz}.

Now, it has been shown that, if this anomaly is real, it implies NP in
$\bs$-$\bsbar$ mixing. Such NP effects can appear in $M_{12}^s$ and/or
$\Gamma_{12}^s$. In fact, it has been argued that NP in
$\Gamma_{12}^s$ should be considered as the main explanation for the
above result \cite{Bobeth}.

In the SM, the dominant contribution to $\Gamma_{12}^s$ is $\bscc$.
Significant NP contributions, i.e.\ comparable to that of the SM, can
come mainly from $\bstautau$. This is (in principle) straightforward
to detect -- if ${\cal B}(\bs \to \tau^+ \tau^-)$ is observed to be at
the percent level, this will be a clear indication of NP (in the SM,
${\cal B}(\bs \to \tau^+ \tau^-) = 7.9 \times 10^{-7}$). Thus, this is
one decay that LHCb should try to measure.

\section{\boldmath $\bs\to V_1 V_2$ Decays}

$\bs\to V_1 V_2$ is really 3 decays. Being vector mesons, $V_1$ and
$V_2$ can have relative orbital angular momentum $l=0$, 1 or 2 ($s$,
$p$ or $d$ wave). This is taken into account by decomposing the decay
amplitude into components in which the polarizations of the
final-state vector mesons are either longitudinal ($A_0$), or
transverse to their directions of motion and parallel ($A_\|$) or
perpendicular ($A_\perp$) to one another.

{\bf\boldmath (1) $\fT$, $\fL$:} Naively, one expects $\fT \ll \fL$,
where $\fT$ ($\fL$) is the fraction of transverse (longitudinal)
decays in $B\to V_1 V_2$.  However, it was observed that $\fTfL \simeq
1$ in $B\to\phi K^*$ \cite{phiK*}.  One explanation of this
``polarization puzzle'' is that the $1/m_B$ penguin-annihilation (PA)
contributions are important \cite{Kagan}.  PA can be sizeable within
QCD factorization (QCDf).

There are two penguin decay pairs whose amplitudes are the same under
flavour SU(3), and for which there is a good estimate of SU(3)
breaking within QCDf: ($\bs \to \phi \phi, \bd \to \phi K^{0*}$) and
($\bs \to \phi {\bar K}^{0*}, \bd \to {\bar K}^{0*} K^{0*}$)
\cite{DMLNS}. Given the polarization in the $\bd$ decay, can predict
the polarization in the $\bs$ decay, and thus test PA.

This has been partially done -- $\bs \to \phi \phi$ has been measured:
\bea
{\rm predict:} ~~~~~ \frac{\fT(\bs \to \phi \phi)}{\fT(\bd \to \phi K^{0*})} & = & 1.36 \pm 0.59 ~, \nn\\ 
{\rm expt:} ~~~~~ \frac{\fT(\bs \to \phi \phi)}{\fT(\bd \to \phi K^{0*})} & = & 1.25 \pm 0.11 ~.
\eea
The theoretical error is large, but there is reasonable agreement.

{\bf (2) Triple Product (TP):} In the $B$ rest frame, the TP takes the
form ${\vec q} \cdot ({\vec\varepsilon}_1 \times
{\vec\varepsilon}_2)$, where ${\vec q}$ is the difference of the two
final momenta, and ${\vec\varepsilon}_1$ and ${\vec\varepsilon}_2$ are
the polarizations of $V_1$ and $V_2$.  The TP is odd under both P and
T, and thus constitutes a potential signal of CPV. There are two TP's:
$A_T^{(1)} \propto {\rm Im}(A_\perp A_0^*)$ and ${A_T^{(2)} \propto
  \rm Im}(A_\perp A_\|^*)$.

The statement that ``TP's are a signal of CP violation'' is not quite
accurate.  The $A_i$ ($i=0,\|,\perp$) possess both weak (CP-odd) and
strong (CP-even) phases. Thus, ${\rm Im}(A_\perp A_0^*)$ and ${\rm
  Im}(A_\perp A_\|^*)$ can both be nonzero even if the weak phases
vanish. In order to obtain a true signal of CP violation, one has to
compare the $B$ and ${\bar B}$ decays.

The TP's for the ${\bar B}$ decay are $-{\rm Im}({\bar A}_\perp {\bar
  A}_0^*)$ and $-{\rm Im}({\bar A}_\perp {\bar A}_\|^*)$, in which
${\bar A}_0$, ${\bar A}_\|$, and ${\bar A}_\perp$ are equal to $A_0$,
$A_\|$, and $A_\perp$, respectively, but with weak phases of opposite
sign. $A_\perp$ is pure $p$ wave ($l=1$), and so the additional minus
sign is generated when CP is applied and $A_\perp \to{\bar A}_\perp$.
The true (CP-violating) TP's are then given by $\frac12[{\rm
    Im}(A_\perp A_0^*) + {\rm Im}({\bar A}_\perp {\bar A}_0^*)]$ and
$\frac12[{\rm Im}(A_\perp A_\|^*) + {\rm Im}({\bar A}_\perp {\bar
    A}_\|^*)]$. 

Now, CPV requires the interference of two amplitudes. The common way
to look for CPV is via a nonzero rate difference between a decay and
its CP-conjugate decay (direct CPV). The direct CP asymmetry is
proportional to $\sin\phi \sin\delta$, where $\phi$ and $\delta$ are
the relative weak and strong phases of the two amplitudes. That is,
direct CPV requires a nonzero strong-phase difference. On the other
hand, the true TP is proportional to $\sin\phi \cos\delta$, so no
strong-phase difference is necessary. This helps in the search for
NP. Also, in the SM, true TP's are generally small (or zero)
\cite{DLTPs}, so that TP's are a good way to find NP.

CDF and LHCb have measured the true TP asymmetries in $\bs
\to\phi\phi$ \cite{CDFLHCbTPs}:
\bea
A_u~(\perp\|) & = & -0.007 \pm 0.064~({\rm stat}) \pm 0.018~({\rm syst}) ~~~~~ {\hbox{CDF}} ~, \nn\\
              & = & -0.064 \pm 0.057~({\rm stat}) \pm 0.014~({\rm syst}) ~~~~~ {\hbox{LHCb}} ~, \nn\\
A_v~(\perp 0) & = & -0.120 \pm 0.064~({\rm stat}) \pm 0.016~({\rm syst}) ~~~~~ {\hbox{CDF}} ~, \nn\\
              & = & -0.070 \pm 0.057~({\rm stat}) \pm 0.014~({\rm syst}) ~~~~~ {\hbox{LHCb}} ~.
\eea
These agree with the SM prediction ($A_u = A_v = 0$).

There are also fake (CP-conserving) TP's, due only to the strong
phases of the $A_i$'s. These are given by $\frac12[{\rm Im}(A_\perp
  A_0^*) - {\rm Im}({\bar A}_\perp {\bar A}_0^*)]$ and $\frac12[{\rm
    Im}(A_\perp A_\|^*) - {\rm Im}({\bar A}_\perp {\bar A}_\|^*)]$.
In the SM, certain fake TP's are very small \cite{fakeTPs}. This
implies that one can partially distinguish the SM from NP through the
measurement of the fake $A_T^{(2)}$ TP. This applies to $\BKstar$ and
$\bs\to\phi\phi$.

\section{Measuring U-spin/SU(3) Breaking}

Consider charmless $\btod$ and $\btos$ decays whose amplitudes are
equal under U spin ($d \leftrightarrow s$). There are four
observables: the CP-averaged $\btod$ and $\btos$ decay rates $B_d$ and
$B_s$, and the direct CP asymmetries $A_d$ and $A_s$. In the U-spin
limit, $X=1$, where $X \equiv -(A_s/A_d) (B_s/B_d)$.  Thus, by
measuring the four observables, and computing the deviation of $X$
from 1, one can {\it measure} U-spin breaking \cite{ImLon}.

This can be applied to 4 decay pairs involving $\bs$ decays: $\bd \to
\pi^+ \pi^-$ and $\bs \to K^+ K^-$, $\bs \to \pi^+ K^-$ and $\bd \to
\pi^- K^+$, $\bd \to K^0 \kbar$ and $\bs \to \kbar K^0 $, $\bd \to K^+
K^-$ and $\bs \to \pi^+ \pi^-$.  The first (second) decay is $\btod$
($\btos$).

If one neglects annihilation- and exchange-type diagrams, there are 12
additional pairs of decays to which this analysis can be
applied. These are not related by U spin, but are instead related by
flavour SU(3).

\section{Conclusions}

In the past, there were a number of hints of NP in some $B$ decays,
usually in $\btos$ transitions. Unfortunately, with recent LHCb
measurements, most of these have gone away. This suggests that, if NP
is present, very large signals are unlikely.

Still, the LHC has the precision to detect small deviations from the SM
predictions. To this end, it is best to make measurements of as many
different processes as possible. In this talk, I have mentioned a
number of different possibilities (some of which have already been
measured).  Hopefully, when these measurements are made, we will see a
sign of NP.

\bigskip
This work was financially supported by NSERC of Canada.

\end{document}